# ICU Patient Deterioration Prediction: A Data-Mining Approach

Noura AlNuaimi, Mohammad M Masud and Farhan Mohammed

College of Information Technology, United Arab Emirates University, Al-Ain, UAE
{noura.alnuaimi, m.masud, 200835338}@uaeu.ac.ae

## ABSTRACT

*A huge amount of medical data is generated every day, which presents a challenge in analysing these data. The obvious solution to this challenge is to reduce the amount of data without information loss. Dimension reduction is considered the most popular approach for reducing data size and also to reduce noise and redundancies in data. In this paper, we investigate the effect of feature selection in improving the prediction of patient deterioration in ICUs. We consider lab tests as features. Thus, choosing a subset of features would mean choosing the most important lab tests to perform. If the number of tests can be reduced by identifying the most important tests, then we could also identify the redundant tests. By omitting the redundant tests, observation time could be reduced and early treatment could be provided to avoid the risk. Additionally, unnecessary monetary cost would be avoided. Our approach uses state-of-the-art feature selection for predicting ICU patient deterioration using the medical lab results. We apply our technique on the publicly available MIMIC-II database and show the effectiveness of the feature selection. We also provide a detailed analysis of the best features identified by our approach.*

## 1. INTRODUCTION

Healthcare is changing from traditional medical practice to modern evidence-based healthcare. Evidence is based on patient data, which are collected from different resources like electronic health record (EHR) systems, monitoring devices and sensors [1]. One specific example of these technological advances is the observation and monitoring technologies for intensive care unit (ICU) patients. Currently, the data generated in the process of medical care ICUs are huge, complex and unstructured. Such data can be called big data due to their complexity, large size and difficulty to process in real-time [2]. However, these data could be used with the help of intelligent systems, such as big data analytics and decision support systems, to determine which patients are at an increased risk of death. This could support making the right decision to enhance the efficiency, accuracy and timeliness of clinical decision making in the ICU.

Reducing the amount of data without losing information is a great challenge. Dimension reduction would be the first solution to eliminate duplicate, useless and irrelevant features. In this paper, our goal is to propose an efficient mining technique to reduce the observation time in ICUs by predicting patient deterioration in its early stages through big data analytics. Our proposed technique has several contributions. First, we use the lab test results to predict patient deterioration. To the best of our knowledge, this is the first work that primarily uses medical lab tests to predict patient deterioration. Lab test results have a crucial role in medical decision making. Second, we identify most important medical lab tests using state-of-the-art feature-selection techniques without using any informed domain knowledge. Finally, our approach helps reduce redundant medical lab tests. Thus, healthcare professionals could focus on the most important lab tests to assist them, which would save not only costs but also valuable time in recovering the patient from a critical condition.

The paper is organised as follows. Section 2 presents the related work of predicting ICU death, Section 3 gives background on data mining and big data analytics, Section 4 illustrates our proposed approach, Section 5 summarises the MIMIC II dataset, Section 6 illustrates the experiment's work, Section 7 discusses the findings, and finally, the conclusion of this research is presented in Section 8.

## 2. LITERATURE REVIEW

This section reviews related works for predicting ICU death or the deterioration of ICU patients. We highlight some similarities and differences between some of the related works and the proposed work.

In [3], the authors developed an integrated data-mining approach to give early deterioration warnings for patients under real-time monitoring in the ICU and real-time data sensing (RDS). They synthesised a large feature set that included first- and second-order time-series features, detrended fluctuation analysis (DFA), spectral analysis, approximative entropy and cross-signal features. Then, they systematically applied and evaluated a series of established data-mining methods, including forward feature selection, linear and nonlinear classification algorithms, and exploratory under sampling for class imbalance. In our work, we are using the same dataset. However, we are using only the medical lab tests. Also, in our approach, we depend on feature selection to reduce the size of the dataset.

A health-data search engine was developed in [4] that supported predictions based on the summarised clusters patient types which claimed that it was better than predictions based on the non-summarised original data. In our work, we use only the medical lab tests, and we attempt to highlight the most important medical labs.

Liu et al. [4] investigated the critical feature size dimension. In their work, an ad hoc heuristic method based on feature-ranking algorithms was used to perform the experiment on six datasets. They found that the heuristic method is useful in finding the critical feature dimension for large datasets. In our work, we also use the ranking to rank the most useful features. However, we attempt to investigate the percentage of selected features that would be enough to have moderate model accuracy.

A survey of feature selection is presented in [6]. The authors presented a basic taxonomy of feature-selection techniques and discussed their use, variety and potential in a number of common and upcoming bioinformatics applications.

Cismondi et al. [5] proposed reducing unnecessary lab testing in the ICU. They applied artificial intelligence to study the predictability of future lab test results for gastrointestinal bleeding. This work is the closest work to our research; they have the same objective of reducing unnecessary lab tests. However, they only focus on gastrointestinal bleeding. In our work, we are targeting all cases in the ICUs.

## 3. BACKGROUND ON DATA MINING AND BIG DATA ANALYTICS

Healthcare, like other sectors, is facing the need for analysing large amounts of information, otherwise known as big data, which has become a major driver of innovation and success. Big data has potential to support a wide range of medical and healthcare functions, including clinical decision support [2].

Data mining is the analysis step of knowledge discovery. It is about the 'extraction of interesting (non-trivial, implicit, previously unknown, and potentially useful) patterns or knowledge from huge amount of data [10]'. When mining massive datasets, two of the most common, important and immediate problems are sampling and feature selection. Appropriate sampling and feature selection contribute to reducing the size of the dataset while obtaining satisfactory results in model building [4].

## 3.1. Feature Selection

In machine learning, feature selection or attribute selection is the process of selecting a subset of relevant features (variables, predictors) for use in model construction. Feature selection techniques are used (a) to avoid overfitting and improve model performance, i.e. predict performance in the case of supervised classification and better cluster detection in the case of clustering, (b) to provide faster and more cost-effective models and (c) to gain deeper insight into the underlying processes that generated the data. In the context of classification, feature selection techniques can be organized into three categories, depending on how they perform the feature selection search to build the classification model: filter methods, wrapper methods and embedded methods, presented in table 1 [6] [7]:

1) Filter Methods are based on applying a statistical measure to assign a scoring to each feature. Then, features are ranked by score and either selected or removed from the dataset. The methods are often univariate and consider the feature independently or with regard to the dependent variable.
2) Wrapper Methods are based on the selection of a set of features as a search problem, where different combinations are prepared, evaluated and compared to other combinations. A predictive model is used to evaluate a combination of features and assign a score based on model accuracy.
3) Embedded Methods are based on learning which features most contribute to the accuracy of the model while the model is being created.

Table 1: Feature selection categories.

| Model Search | Advantages | Disadvantages |
|---|---|---|
| Filter | Fast<br>Scalable<br>Independent of the classifier | Ignores feature dependencies<br>Ignores interaction with the classifier |
| Wrapper | Simple<br>Interacts with the classifier<br>Models feature decencies<br>Less computational | Risk for overfitting<br>More prone than randomized algorithms<br>Classifier-dependent selection |
| Embedded | Interacts with the classifier<br>More computational<br>Models feature dependencies | Classifier-dependent selection |

## 3.2. Data Classification Techniques

Classification is a pattern-recognition task that has applications in a broad range of fields. It requires the construction of a model that approximates the relationship between input features and output categories [8]. Some of the most popular techniques are discussed here in brief, all of which are used in our work.

1) The Naïve Bayes classifier is based on applying Bayes' theorem with strong independence assumptions between the features. As one of its main features, the Naïve Bayes classifier is easy to implement because it requires a small amount of training data in order to estimate the parameters, and good results can be found in most cases. However, it has class conditional independence, meaning it causes losses of accuracy and dependency [9].
2) Sequential minimal optimization (SMO) is an algorithm for efficiently solving the optimization problem which arises during the training of support vector machines [10]. The amount of memory required for SMO is linear in the training set size, which allows SMO to handle very large training sets [11].

3) The ZeroR classifier simply predicts the majority category, which relies on the target and ignores all predictors. Although there is no predictability power in ZeroR, it is useful for determining a baseline performance as a benchmark for other classification methods [10].
4) A decision tree (J48) is a fast algorithm to train and generally gives good results. Its output is human readable, therefore one can see if it makes sense. It has tree visualizers to aid understanding. It is among the most used data mining algorithms. The decision tree partitions the input space of a data set into mutually exclusive regions, each of which is assigned a label, a value or an action to characterize its data points [10].
5) A RandomForest is a combination of tree predictors such that each tree depends on the values of a random vector sampled independently and with the same distribution for all trees in the forest [12].

## 4. PROPOSED APPROACH

In this section we introduce our approach for the Big Data mining technique for predicting ICU patient deterioration. Figure 1 shows the architecture of the proposed technique.

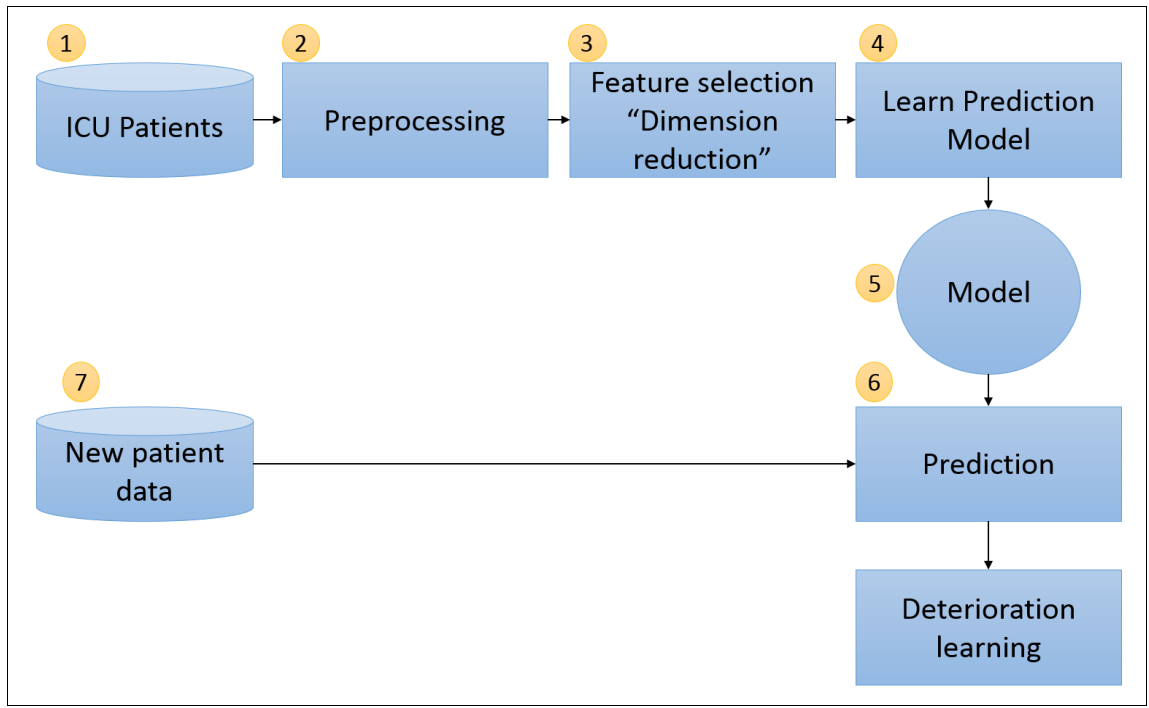

Figure 1: Architecture of the proposed approach.

The data are collected from the database of ICU patients (step 1). Then the data are integrated, cleaned and relevant features are extracted (step 2). After that, feature selection or dimensionality reduction techniques are applied to obtain the best set of features and reduce the data dimension (step 3). Then the prediction model is learned using a machine learning approach (step 4). When a new patient is admitted to the CPU, the patient's data are collected incrementally (step 5). The patient data are evaluated by the prediction model (step 6) to predict the possibility of deterioration of the patient, and warnings are generated accordingly. Each of these steps is summarized here, and more details of the dataset are given in Section 5.

1) ICU Patient Data: The details of the data and the collection process are discussed in Section 5.

2) Preprocessing: At the preprocessing stage, we used two different datasets. These datasets were generated from a Labevents table. The first dataset contained the average value of applied medical tests, and the second contained the total number of times for each test was applied.
3) Feature Selection / Dimension Reduction: attribute selection is the process of selecting a subset of relevant features (variables, predictors) for use in model construction. The goal here is to reduce the attributes so medical professional can identify the most important medical lab tests used by reducing the redundant tests. In our work, we select filter methods because they are moderately robust against the overfitting problem, as follows:
    a. Attribute evaluator: InfoGrainAttributeEval
    b. Search method: Ranker
    c. Attribute selection mode: use full training set
4) Learning: In our experiment we use a classification technique and five of the most popular classifier techniques: Naïve Bayes classifier, Support vector machine (SVM), ZeroR classifier, decision tree (J48) and RandomForest. We use different types of machine learning order to avoid random results.
5) Model: The developed model aims to predict ICU patient deterioration by mining lab test results. Thus, observation time can be reduced in the ICUs and more actions can be taken in the early stages.
6) New patient data: When a new patient is admitted to the ICU, all his information is stored in the database. Some of these are incremental, such as vital sign readings, lab test results, medication events etc. The data of the patient again go through the preprocessing and feature extraction phases before they can be applied to the model.
7) Prediction: After each new test result, medication event, etc., the patient data are preprocessed and features are extracted to supply to the prediction model. The model predicts the probability of deterioration for the patient. This probability may change when new data (e.g. more test results) are accumulated and applied to the model. When the deterioration probability reaches a certain threshold specified by the healthcare providers, a warning is generated. This would help the healthcare providers to take proactive measures to save the patient from getting into a critical or fatal condition.

## 5. MIMIC II DATABASE

The MIMIC-II database is part of the Multiparameter Intelligent Monitoring in Intensive Care project funded by the National Institute of Biomedical Imaging and Bioengineering at the Laboratory of Computational Physiology at MIT, which was collected from 2001 to 2008 and represents 26,870 adult hospital admissions. In our work, we use MIMIC-II version 2.6 because is more stable than the newer version 3, which is still in the beta phase and needs further work of cleaning, optimizing and testing. MIMIC-II consists of two major components: clinical data and physiological waveforms.

The MIMIC dataset has three main features: (1) it is public; (2) it has a diverse and very large population of ICU patients; and (3) it contains high temporal resolution data, including lab results, electronic documentation, and bedside monitor trends and waveforms[13]. Several works have used the MIMIC dataset, such as [14], [15] and [16].

In our work, we focus on the clinical data, the LABEVENTS and LABITEMS tables. The Labevents table contains data of each patient's ICU stay, as presented in table 2, and table 3 contains descriptions of the lab events. Considering medical lab choice was done because we wanted to investigate the relationship between medical lab tests and patient deterioration so we could identify which medical tests have a major effect on clinical decision making. For example, the following information is about a patient who was staying at the ICU and was given a medical test. The following information was recorded at that time:
- Subject_ID: 2
- Hadm_ID: 25967

- IcuStay_ID: 3
- ItemID: 50468
- Charttime: 6/15/2806 21:48
- Value: 0.1
- ValueNum: 0.1
- Flag: abnormal
- ValueUOM: K/uL

Table 2: Labevents Table Description

| Name | Type | Null | Comment |
| --- | --- | --- | --- |
| SUBJECT_ID | NUMBER(7) | N | Foreign key, referring to a unique patient identifier |
| HADM_ID | NUMBER(7) | Y | Foreign key, referring to the hospital admission ID of the patient |
| ICUSTAY_ID | NUMBER(7) | Y | ICU stay ID |
| ITEMID | NUMBER(7) | N | Foreign key, referring to an identifier for the laboratory test name |
| CHARTTIME | TIMESTAMP(6) WITH TIME ZONE | N | The date and time of the test |
| VALUE | VARCHAR2(100) | Y | The result value of the laboratory test |
| VALUENUM | NUMBER(38) | Y | The numeric representation of the laboratory test if the result was numeric |
| FLAG | VARCHAR2(10) | Y | Flag or annotation on the lab result to compare the lab result with the previous or next result |
| VALUEUOM | VARCHAR2(10) | Y | The units of measurement for the lab result value |

Table 3: Labitems Table

| Name | Type | Null | Comment |
| --- | --- | --- | --- |
| ITEMID | NUMBER(7) | N | Table record unique identifier, the lab item ID |
| TEST_NAME | VARCHAR2(50) | N | The name of the lab test performed |
| FLUID | VARCHAR2(50) | N | The fluid on which the test was performed |
| CATEGORY | VARCHAR2(50) | N | Item category |
| LOINC_CODE | VARCHAR2(7) | Y | LOINC code for lab item |
| LOINC_DESCRIPTION | VARCHAR2(100) | Y | LOINC description for lab item |

## 6. EXPERIMENTS

We conducted four experiments to fulfil the different approaches to reach our goal of predicting ICU patient deterioration by mining lab test results. In each experiment, a different dataset resulted from pre-processing the MIMIC II v2.6 database.

### 6.1. Experiment 1: Building a Baseline of the Medical Lab Tests Average

1) Experiment Goal: The goal of this experiment was to investigate the effect of lab testing on predicting patient deterioration. Usually, medical professionals compare the result of the lab test with a reference range [17]. If the value is not within this range, the patient may face fatal consequences. Thus, the patient is kept under observation and the test is repeated again during a specific period. In our experiment, we investigated the average value of the same repeated test and, more precisely, how the average value of lab results could assist medical professionals in evaluating patient status.
Since we dealt with real cases, the only way to assess the quality and characteristics of a data mining model was through the final status of the patient, i.e. whether the patient survived or not. Thus, our evaluation criterion was how accurately our approach could predict whether the patient died or not.

2) Building the Dataset: The dataset was constructed by taking the average test result of each patient for each kind of test and make it one attribute. Thus one patient would be represented as one instance having 700 attributes, one for each test. If a test was not done, then the value of that attribute would be 0. For example, the first patient record in the dataset would look like this:
P_ID    Avg1    Avg2    .....    Avg700    Dead/Alive
1       5.3     10               0         D

3) Pre-processing: After building the dataset, some values could not be reported because they were in text format. We used default values for these types of data. The total number of attributes was 619 with 2900 instances.

4) Base learners: In our experiment we used five classification algorithms to construct the model, namely NaiveBayes, SMO, ZeroR, J48 and RandomForest.

Table 4: Experiment 1 results.

| Algorithm | Learning Machine | Detailed Accuracy | | | |
|---|---|---|---|---|---|
| | | Accuracy | Precision | Recall | F-Measure |
| Bayes | NavieBayes | 42.96% | 0.672 | 0.430 | 0.404 |
| Functions | SMO | 76.86 % | 0.759 | 0.769 | 0.762 |
| Rule | ZeroR | 70.24 % | 0.493 | 0.702 | 0.580 |
| Tree | J48 | 75.27% | 0.749 | 0.753 | 0.751 |
| Tree | RandomForest | 77.58 % | 0.765 | 0.776 | 0.762 |

5) Evaluation: For a performance measurement, we did a 10-fold cross-validation of the dataset, and the confusion matrix was obtained to estimate four measures: accuracy, sensitivity, specificity and F-measure. As a result, RandomForest had the highest accuracy of 77.58%, followed by SMO with 76.86%, J48 with 75.27%, ZeroR with 70.24% and NavieBayes with 42.96%, as shown in Table 4. RandomForest and SMO have the same F-measures. The reason for the best performance by RandomForest is that it works relatively well when used with high-dimensional data with a redundant/noisy set of features [12].

**6.2. Experiment 2: Average Medical Lab Tests Feature Selection**

1) Experiment Goal: The goal of this experiment was to study the relationship between feature selection and classification accuracy. Feature selection is one of the dimensionality reduction techniques for reducing the attribute space of a feature set. More precisely, it determines how many features should be enough to give moderate accuracy.
2) Building the Dataset: In this experiment we used the same dataset that we used in experiment 1.
3) Pre-processing: In this experiment we built ten datasets depending on the number of selected features. We start with the first dataset, which contained only 10% of the total attributes. Then each time, we increased the total feature selections by 10%. For example, dataset 1 contains 10% of the total attributes, dataset 2 contains 20% of the total attributes, dataset 3 contains 30% of the total attributes and so on till dataset 10 contains all 100% of the total attributes.
   For feature selection, we use supervised.attribute. InfoGainAttributeEval from WEKA. This filter is a wrapper for the Weka class that computes the information gain on a class [18].
   - Attribute Subset Evaluator: InfoGainAttributeEval
   - Search Method: Ranker.
   - Evaluation mode: evaluate all training data

4) Base learner: After generating all of the reduced datasets, we use the J48 algorithm to construct a model.

Table 5: Experiment 2 Feature selection.

| % of Features Selected | # of Features Selected | Detailed Accuracy | | |
|---|---|---|---|---|
| | | Accuracy | Number of leaves | Size of the Tree |
| 10% | 62 | 75.10% | 200 | 399 |
| 20% | 124 | 73.59% | 201 | 401 |
| 30% | 186 | 75.10% | 185 | 369 |
| 40% | 248 | 74.93% | 179 | 357 |
| 50% | 310 | 75.17% | 189 | 377 |
| 60% | 371 | 74.79% | 187 | 373 |
| 70% | 433 | 75.00% | 189 | 377 |
| 80% | 495 | 75.31% | 184 | 367 |
| 90% | 557 | 74.97% | 183 | 365 |
| 100% | 619 | 74.86% | 184 | 367 |

5) Evaluation: For each reduced dataset, we applied 10-fold cross-validation for evaluating the accuracy. Table V shows the results in numbers, and Figure 2 shows them as a chart. The results indicate that taking only the most related 10% of the total features can give a 75.10% accurate result, which is comparable to the accuracy of the full feature set. This indicates that not all of the features are required to get the highest accuracy. However, there are some fluctuations, such as at 20%, the accuracy drops a little. We conclude that selecting 50 to 80% of the attributes should give moderately satisfying accuracy.

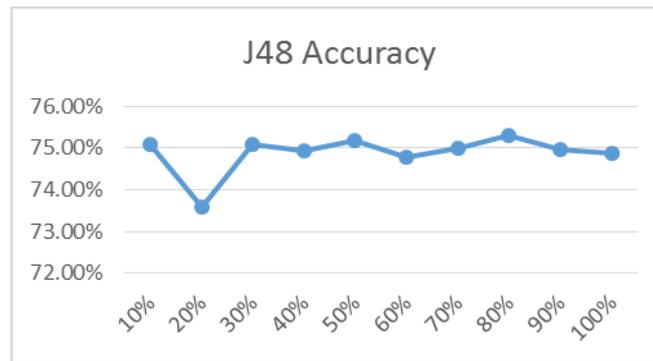

Figure 2: Average datasets accuracy.

### 6.3. Experiment 3: Building a Baseline for the Total Number of Medical Lab Tests

1) Experiment Goal: The goal of this experiment was to investigate the effect of the total number of lab tests conducted on predicting patient deterioration. Usually, medical professionals keep requesting the same medical test over a brief period to compare the result with a reference range [17]. If the value is not within the range, it means the patient may be in danger, so the test is repeated again and again. Our goal was to predict at what total number a medical professional should start immediate action and, more precisely, how the total number of medical lab tests could assist the medical professional in evaluating the patient's status.
2) Building the Dataset: The dataset was built by taking the total number of tests taken for each patient for each type of test and make it one attribute. Then one patient would be represented as one instance having 700 attributes, one for each test. If a test was not done, then the value of that attribute would be 0. For example, the dataset would look like this:
P_ID      Count1 Count2 …    Count700      Dead/Alive
1          5           0                      1               D

3) Pre-processing: The dataset was randomized first, then two datasets were generated, Count_Training_Validation_Dataset and Count_testing_Dataset. This step was repeated ten times because we used randomization to distribute the instances between the two datasets.
4) Base learners: Five learning algorithms were used to build the model, namely NaiveBayes, SMO, ZeroR, J48 and RandomForest.

Table 6: Experiment 3 results.

| Algorithm | Learning Machine | Detailed Accuracy | | | |
|---|---|---|---|---|---|
| | | Accuracy | Precision | Recall | F-Measure |
| Bayes | NavieBayes | 73.66% | 0.718 | 0.737 | 0.713 |
| Funtions | SMO | 75.44% | 0.739 | 0.755 | 0.723 |
| Rule | ZeroR | 70.46% | 0.497 | 0.705 | 0.583 |
| Tree | J48 | 73.16% | 0.728 | 0.732 | 0.692 |
| Tree | RandomForest | 75.73% | 0.742 | 0.757 | 0.739 |

Table 7: Experiment 3 Results.

| Algorithm | Learning Machine | Detailed Accuracy | | | |
|---|---|---|---|---|---|
| | | Accuracy | Precision | Recall | F-Measure |
| Bayes | NavieBayes | 73.48% | 0.716 | 0.735 | 0.711 |
| Funtions | SMO | 74.85% | 0.737 | 0.749 | 0.716 |
| Rule | ZeroR | 69.72% | 0.486 | 0.697 | 0.573 |
| Tree | J48 | 72.44% | 0.722 | 0.724 | 0.723 |
| Tree | RandomForest | 75.30% | 0.739 | 0.753 | 0.736 |

5) Evaluation: The training data were first used to build the model and then evaluated using a percentage split via test data. For a performance measurement, the confusion matrix was obtained to estimate four measures: accuracy, sensitivity, specificity and F-measure. Table 6 shows that SMO and RandomForest have almost equal levels of accuracy, around 75%. Even after testing the model with the test datasets, SMO and RandomForest still have the highest accuracy among the other techniques. The reason for this higher accuracy is that the amount of memory required for SMO is linear in the training set size, which allows SMO to handle very large training sets [11].

### 6.4. Experiment 4: Feature Selection for Total Number of Medical Lab Tests

1) Experiment Goal: The goal of this experiment was to study the relationship between feature selection and classification accuracy. Feature selection is one of the dimensionality reduction techniques for reducing the attribute space of a feature set. More precisely, it measures how many features should be enough to give moderate accuracy.
2) Building the Dataset: In this experiment we used a count dataset.
3) Pre-processing: In the pre-processing step, we built ten datasets depending on the number of selected features. The first dataset contained only 10% of the total attributes. Then we increased the total feature selections by 10% with each new dataset. For example, dataset 1 contained 10% of the total attributes, dataset 2 contained 20% of the total attributes, dataset 3 contained 30% of the total attributes and so on till dataset 10 contained all 100% of the total attributes.
4) For feature selection, we used supervised.attribute. InfoGainAttributeEval from WEKA. This filter is a wrapper for the Weka class that computes the information gain on a class [18].
   - Attribute Subset Evaluator: InfoGainAttributeEval
   - Search Method: Ranker.
   - Evaluation mode: evaluate on all training data

5) Base learner: After generating all reduced datasets, we used the J48 algorithm as a base learner.

Table 8: Experiment 4 Results.

| % of Features Selection | # of Features Selection | Detailed Accuracy | | |
|---|---|---|---|---|
| | | Accuracy | Number of leaves | Size of the Tree |
| 10% | 62 | 71.45% | 237 | 473 |
| 20% | 124 | 73.90% | 250 | 499 |
| 30% | 186 | 73.55% | 247 | 493 |
| 40% | 248 | 72.79% | 252 | 503 |
| 50% | 310 | 73.41% | 252 | 503 |
| 60% | 371 | 73.66% | 254 | 507 |
| 70% | 433 | 74.24% | 254 | 507 |
| 80% | 495 | 74.10% | 254 | 507 |
| 90% | 557 | 74.14% | 265 | 529 |
| 100% | 619 | 73.59% | 259 | 517 |

6) Evaluation: Each feature-reduced dataset went through a 10-fold cross-validation for evaluation. Figure 3 shows the accuracy of all count datasets. The detail values are also reported in Table 4. From the results we observe that selecting 60 to 70% of the attributes gives the highest accuracy. This also concludes that all features (i.e., lab tests) may not be necessary to attain a highly accurate prediction of patient deterioration.

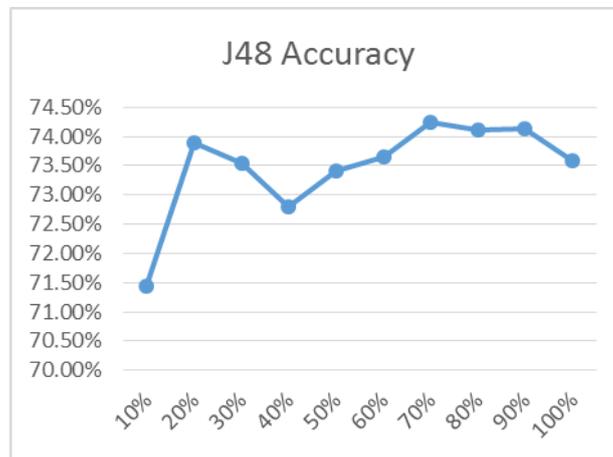

Figure 3: Count dataset accuracy.

## 7. DISCUSSION

It should be noted that the feature selections were done without any domain knowledge and without any intervention from medical experts. However, in the analysis we would like to emphasize the merit of feature selection in choosing the best tests, which could be further verified and confirmed by a medical expert.

First we compare the selected features selected from the two datasets, namely the average dataset and the count dataset. Table 9 shows the 10 best features chosen by the two approaches and highlights the common lab tests between the two approaches (i.e. using the average of tests and count of tests). Table 10 shows more details about the common tests.

Table 9: Final Results.

|  | Detailed Accuracy | |
|---|---|---|
|  | **Average Dataset** | **Count Dataset** |
| Best ranked 10 from the 10% of selected features | 50177<br>50090<br>50060<br>50399<br>50386<br>50440<br>50408<br>50439<br>50112<br>50383 | 50148<br>50112<br>50140<br>50399<br>50177<br>50439<br>50090<br>50440<br>50079<br>50068 |

Table 10: Medical Lab Test Details.

|  | Detailed Description | | | | |
|---|---|---|---|---|---|
|  | **Test_Name** | **Fluid** | **Category** | **LOINC_Code** | **LOINC_Desc** |
| 50177 | UREA N | BLOOD | CHEMISTRY | 3094-0 | Urea nitrogen [mass/volume] in serum or plasma |
| 50090 | CREAT | BLOOD | CHEMISTRY | 2160-0 | Creatinine [mass/volume] in serum or plasma |
| 50399 | INR(PT) | BLOOD | HEMATOLOGY | 34714-6 | INR in blood by coagulation assay |

| | Detailed Description | | | | |
|---|---|---|---|---|---|
| | Test_Name | Fluid | Category | LOINC_Code | LOINC_Desc |
| 50440 | PTT | BLOOD | HEMATOLOGY | 3173-2 | Activated partial thromboplastin time (aPTT) in blood by coagulation assay |
| 50439 | PT | BLOOD | HEMATOLOGY | 5964-2 | Prothrombin time (PT) in blood by coagulation assay |
| 50112 | GLUCOSE | BLOOD | CHEMISTRY | 2345-7 | Glucose [mass/volume] in serum or plasma |

LOINC is an abbreviation for logical observation identifiers names and codes. LOINC is clinical terminology important for laboratory test orders and results [19]. ARUP Laboratories [20] is a national clinical and anatomic pathology reference laboratory and a worldwide leader in innovative laboratory research and development. We used their web page and others to clarify more about the medical lab tests in table 10 as follows:

- UREAN (50177): This test is conducted using the patient's blood. This test is recommended to screen for kidney dysfunction in patients with known risk factors (e.g. hypertension, diabetes, obesity, family history of kidney disease). The panel includes albumin, calcium, carbon dioxide, creatinine, chloride, glucose, phosphorous, potassium, sodium and BUN and a calculated anion gap value. Usually, the result is reported within 24 hours [20].
- CREAT (50090): This test is conducted using the patient's blood. It is a screening test to evaluate kidney function [20].
- INR(PT) (50399): This test is conducted using the patient's blood by coagulation assay [13].
- PTT (50440): This test is carried out to answer two main questions: does the patient have antiphospholipid syndrome (APLS), and does the patient have von Willebrand disease? If so, which type? It is carried out by mechanical clot detection [21].
- PT (50439): This test is conducted using the patient's blood by coagulation assay [13].

- GLUCOSE (50112): This test is used to check glucose, which is a common medical analytic measured in blood samples. Eating or fasting prior to taking a blood sample has an effect on the result. Higher than usual glucose levels may be a sign of prediabetes or diabetes mellitus [22].
- The result of the top 10 selected features from the average dataset allows us to build a model using decision tree J48. This model would allow a medical professional to predict the status of a patient in the ICU as follows:

```
50440 <= 20.757143: 1 (772.0/22.0)
50440 > 20.757143
|   50177 <= 25.923077
|   |   50060 <= 0
|   |   |   50112 <= 138.333333
|   |   |   |   50383 <= 28.155556
|   |   |   |   |   50112 <= 110.470588
|   |   |   |   |   |   50399 <= 1.204545: 0 (5.0)
```

For example, if the lab test (name: PTT, ID 50440, LOINC: 3173-2) result value is <= 20.757143, then the probability is very high (772.0/22.0~ 97.2%) that the patient is going to die (class:1). This model has 78.6897% overall accuracy.

## 8. CONCLUSION AND FUTURE WORK

In this paper, we presented our proposed approach to reduce the observation time in the ICU by predicting patient deterioration in its early stages. In our work, we presented experiments 1 and 3 to build a model to predict patient deterioration. Experiments 2 and 4 identified the most important medical lab tests, then highlighted the common tests between the two datasets. The four experiments would help medical professionals to take better decisions in a very short time.

For future work, the authors are planning to carry out more experiments using bigger data. Big data analytics would bring potential benefits to support taking the right decision to enhance the efficiency, accuracy and timeliness of clinical decision making in the ICU.

**Authors**

Noura Al Nuaimi is pursuing a PhD in Information Technology with Dr Mohammad Mehedy Masud at United Arab Emirates University (UAEU). She holds an MSc in Business Administration from Abu Dhabi University and a BSc in Software Engineering from UAEU. Her research interests focus on data mining and knowledge discovery, cloud computing, health information systems, search engines and natural language processing. She has published research papers in *IEEE Computer Society a*nd *IEEE Xplore*.

Dr Mohammad Mehedy Masud is currently an Assistant Professor at the United Arab Emirates University (UAEU). He joined the College of Information Technology at UAEU in spring 2012. He received his PhD from University of Texas at Dallas (UTD) in December 2009. His research interests are in data mining, especially data stream mining and big data mining. He has published more than 30 research papers in journals including *IEEE Transactions on Knowledge and Data Engineering (TKDE), Journal of Knowledge and Information Systems (KAIS), ACM Transactions on Management Information Systems (ACM TMIS)* and peer-reviewed conferences including *IEEE International Conference on Data Mining (ICDM), European Conference on Machine Learning (ECML/PKDD) and Pacific Asia Conference on KDD*. He is the principal inventor of a US patent application and lead author of the book "Data Mining Tools for Malware Detection". Dr Masud has served as a program committee member of several prestigious conferences and has been serving as the official reviewer of several journals, including IEEE TKDE, IEEE TNNLS and DMKD. During his service at the UAEU he has secured several internal and external grants as PI and co-PI.

Farhan Mohammed is a graduate from the College of Information Technology in United Arab Emirates University specializing in Information Technology Management. He obtained his Bachelor's in Management Information Systems from United Arab Emirates University, Al Ain, UAE. He has worked under several professors and published four conference papers and a journal paper for IEEE sponsored conferences. Currently he is working as a research assistant in data mining in the health industry to develop models on health deterioration prediction. His area of interests lies in smart cities, UAVs, data mining, and image and pattern recognition.